\documentclass{emulateapj}
 
\shorttitle{Oblique Orbit of WASP-107b}
\shortauthors{Dai et al.}

\usepackage{amsmath}
\usepackage{epstopdf}
\usepackage{apjfonts}
\usepackage{xspace}
\usepackage{hyperref} 
\usepackage{bm}
\usepackage{graphicx}
\begin{document}


\title{The Oblique Orbit of WASP-107\MakeLowercase{b} from {\it K2} Photometry}


\author{Fei\ Dai\altaffilmark{1,2}, Joshua N.\ Winn\altaffilmark{2}}


\altaffiltext{1}{Department of Physics and Kavli Institute for
  Astrophysics and Space Research, Massachusetts Institute of
  Technology, Cambridge, MA 02139, USA {\tt fd284@mit.edu}}

\altaffiltext{2}{Department of Astrophysical Sciences, Peyton Hall, 4
  Ivy Lane, Princeton, NJ 08540 USA}


\begin{abstract}
 \noindent
  
Observations of nine transits of WASP-107 during the {\it K2} mission
reveal three separate occasions when the planet crossed in front of a
starspot. The data confirm the stellar rotation period to be 17 days
--- approximately three times the planet's orbital period --- and
suggest that large spots persist for at least one full rotation.  If
the star had a low obliquity, at least two additional spot crossings
should have been observed. They were not observed, giving evidence for
a high obliquity. We use a simple geometric model to show that the
obliquity is likely in the range 40-140$^\circ$, i.e., both spin-orbit
alignment and anti-alignment can be ruled out.  WASP-107 thereby joins
the small collection of relatively low-mass stars hosting a giant
planet with a high obliquity. Most such stars have been observed to
have low obliquities; all the exceptions, including WASP-107, involve
planets with relatively wide orbits (``warm Jupiters'', with $a_{\rm
  min}/R_\star \gtrsim 8$). This demonstrates a connection between
stellar obliquity and planet properties, in contradiction to some
theories for obliquity excitation.

\end{abstract}


\section{Introduction}

A star's obliquity is a fundamental geometric property of a planetary
system, and an intriguing piece of the puzzle of planet formation and
orbital evolution \citep[as recently reviewed
  by][]{WinnFabrycky2015}. Many methods have been devised to test for
alignment between stellar rotation and planetary orbital motion
\citep{Queloz2000,Barnes2009,Schlaufman2010,Sanchis-Ojeda2011Wasp,Nutzman2011,Hirano2012,Chaplin2013,Mazeh2015}.
One of these methods, the analysis of starspot crossings, takes
advantage of precise and continuous time-series photometry of stars
with transiting planets, which the {\it Kepler} mission has provided
\citep{Borucki2011}. When a transiting planet crosses in front of a
starspot, the loss of light is briefly reduced, because the hidden
portion of the star has a lower intensity than the surrounding
photosphere.  The detection and timing of these events can sometimes
reveal the stellar obliquity.  One approach is to seek evidence for
multiple crossings of the same spot, which are more likely to occur
for a spin-aligned star than a tilted star \citep[see,
  e.g.,][]{Sanchis-Ojeda2011Wasp,Nutzman2011,Desert2011}.

WASP-107b is a transiting planet discovered by
\citet{Anderson2017}. The host star is a K dwarf of mass
0.69~$M_\star$. The planet has an orbital period of 5.7 days and a
radius of 0.95~$R_{\text{Jup}}$.  Despite this near-Jovian size,
radial-velocity monitoring revealed the planet's mass to be only about
twice that of Neptune ($0.12~M_{\text{Jup}}$). This makes the planet
difficult to classify, and demonstrates that such a relatively low
mass is sufficient to accrete a large gaseous envelope.

WASP-107 happens to be within the field of view of {\it K2} Campaign
10, one of the star fields along the ecliptic that are being monitored
by the {\it Kepler} telescope \citep{Howell2014}. Thanks to a proposal
by \citet{Anderson2017}, WASP-107 was observed with one-minute
sampling (``short cadence''). In this paper we report on our analysis
of the {\it K2} photometry and our assessment of the stellar
obliquity, based on observations of starspot crossings. The following
section describes the data reduction. In Section 3 we refine the basic
transit parameters and identify spot crossings. Section 4 is concerned
with the stellar rotation period, a crucial ingredient in the analysis
of starspot crossings. In Section 5 we present evidence for a high
obliquity by modeling the intersections between the planet's transit
chord and the possible paths of starspots. We discuss the implications
in Section 6.

\section{{\it K2} Photometry}

WASP-107 (or EPIC~228724232) was observed by {\it Kepler} from
2016~Jul~6 to Sep~20 in the short-cadence mode. We disregarded the
first 6 days of data, which were of lower quality because of a
3.5-pixel pointing error that was subsequently corrected.  A further
complication was the loss of Module 4 during this campaign, which
resulted in a 14-day gap in data collection.  Thus the useful data
comprises an interval of about 7 days, followed by the 14-day gap, and
then a continuous interval of nearly 50 days.

\begin{figure*}
\begin{center}
\includegraphics[width = 2.2\columnwidth]{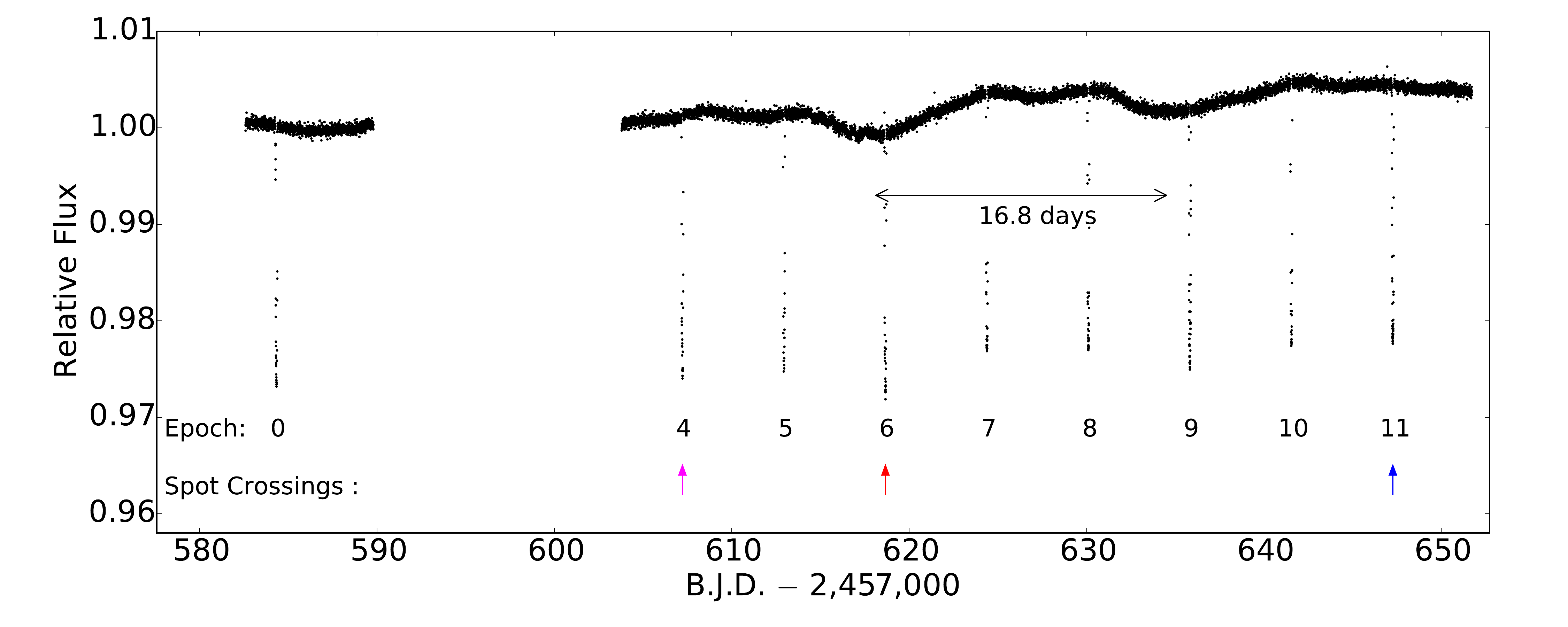}
\caption{ {\bf Corrected {\it K2} photometry of WASP-107.}  Colored arrows
  indicate the transits during which spot-crossing anomalies were
  detected. Outside of transits, the time between the two clearest
  minima is 16.8~days, which agrees with the previously measured
  rotation period.}
\label{fig: lc}
\end{center}
\end{figure*}

We downloaded the pixel files from the Mikulski Archive for Space
Telescopes. To remove the spurious intensity fluctuations caused by
the rolling motion of the spacecraft, we used an approach similar to
that described by \citet{VJ2014}. We considered a circular aperture of
radius 4.5~pixels centered on the brightest pixel, and fitted a 2-d
Gaussian function to the intensity distribution within the aperture.
Then we decorrelated the aperture-summed flux and the $X$ and $Y$
positions of the fitted Gaussian function.  Fig.~\ref{fig: lc} shows
the corrected time series, including stellar variability, transit
signals, and doubtless some residual systematic effects.

\section{Light curve analysis}
\label{lc_analysis}

\citet{Anderson2017} found the stellar rotation period to be nearly
three times as long as the orbital period. If the stellar obliquity
were near $0^\circ$ or $180^\circ$, then the spots would move along
the transit chord as the star rotates. In such a situation, whenever
we see a spot-crossing anomaly during transit epoch $n$, we would also
expect to see one in epochs $n-3$ or $n+3$, as long as the starspot
persists for at least one rotation. We checked for this pattern by
scrutinizing the data obtained during each of the 9 transits.  A
spot-crossing anomaly was obvious to the eye in epoch 4, but none were
visible in epoch 7 (and epoch 1 occurred during the data gap). A
second spot crossing took place in epoch 6, but none were detected in
epochs 3 or 9.  A third possible anomaly, weaker than the others, was
seen in epoch 11 but none were seen in epoch 8. Thus it was
immediately clear that WASP-107 is unlikely to have either a nearly
perfectly prograde or retrograde orbit. A tilt of order $R_p/R_\star$
radians ($\sim$10$^\circ$) is needed for the spots to leave the
transit chord as they rotate across the visible stellar hemisphere.

For quantitative analysis we modeled each transit light curve with the {\tt Batman}
code written by \citet{Kreidberg2015}.  We considered a 7-hour
window around the time of minimum light, and allowed the
out-of-transit flux to be a quadratic function of time, to account for
longer-term stellar variability. The free parameters included the
planet-to-star radius ratio ($R_p/R_\star$); the ratio of stellar
radius to orbital distance ($R_\star/a$); and the impact parameter ($b
\equiv a\cos I/R_\star$). We took the limb-darkening profile to be
quadratic, with both coefficients $u_1$ and $u_2$ as free parameters.
We also took into account the effect of untransited starspots, which
increase the transit depth beyond what it would be in the absence of
spots.  To do so, we introduced an additional parameter $\Delta
F_{\text{spot}}$ specific to each transit, such that
\begin{equation}
  F_{\text{calc,~spot}} = \frac{F_{\text{calc,~no-spot}} - \Delta F_{\text{spot}}}
  {1 - \Delta F_{\text{spot}}}.
\end{equation}
Here, $F_{\text{calc,~no-spot}}$ is the calculated flux using {\tt
  Batman}, and $F_{\text{calc,~spot}}$ is the calculated flux that is
compared to the observed flux.  With this definition,
$F_{\text{calc,~spot}} \equiv 1$ outside of the transits; this is
needed because we normalized the data in this manner.
We adopted the usual $\chi^2$ likelihood function and found
the maximum-likelihood solution using the Levenberg-Marquardt
algorithm as implemented in the {\tt Python} package {\tt lmfit}
\citep{lmfit}.

We modeled the spot-crossing anomalies as Gaussian functions of time:
 \begin{equation}
 \label{eqn:ano}
F_{\text{anom}}(t) = A \exp{\left[-\frac{(t-t_{\text{anom}})^2}{2\sigma_{\text{anom}}^2}\right]}
\end{equation}
 where $A$, $t_{\text{anom}}$ and $\sigma_{\text{anom}}$ represent
 respectively the amplitude, time, and duration of the anomaly.
 To identify anomalies more objectively
 we fitted each transit several times: with no anomalies, one anomaly, two anomalies, and so forth.
 Then we used the Bayesian Information Criterion,
\begin{equation}
\text{BIC}  = 2 \text{log}(L_{\text{max}})+N~\text{log}(M),
\end{equation}
to decide which number of anomalies provides the best fit to the data.
In this equation, $L_{\text{max}}$ is the maximum likelihood, $N$ is
the number of model parameters, and $M$ is the number of data points.
We demanded $\Delta$BIC~$>10$. Only three anomalies passed this
criterion: the same three that were obvious to the
eye. Table~\ref{tab:anomalies} reports their properties, based on the
Monte Carlo Markov Chain (MCMC) code of \citep{emcee}.  The quoted value is
based on the 50\% level of the cumulative posterior distribution, and
the uncertainty interval is based on the 16\% and 84\% levels.

To refine the transit parameters, we excluded data within
$2\sigma_{\text{anom}}$ of the fitted time of each anomaly. Then we
produced a phase-folded, anomaly-free light curve, which was fitted
with another MCMC analysis. Table~\ref{tab:para} reports the results.


\begin{figure*}
\begin{center}
\includegraphics[width = 2.2\columnwidth]{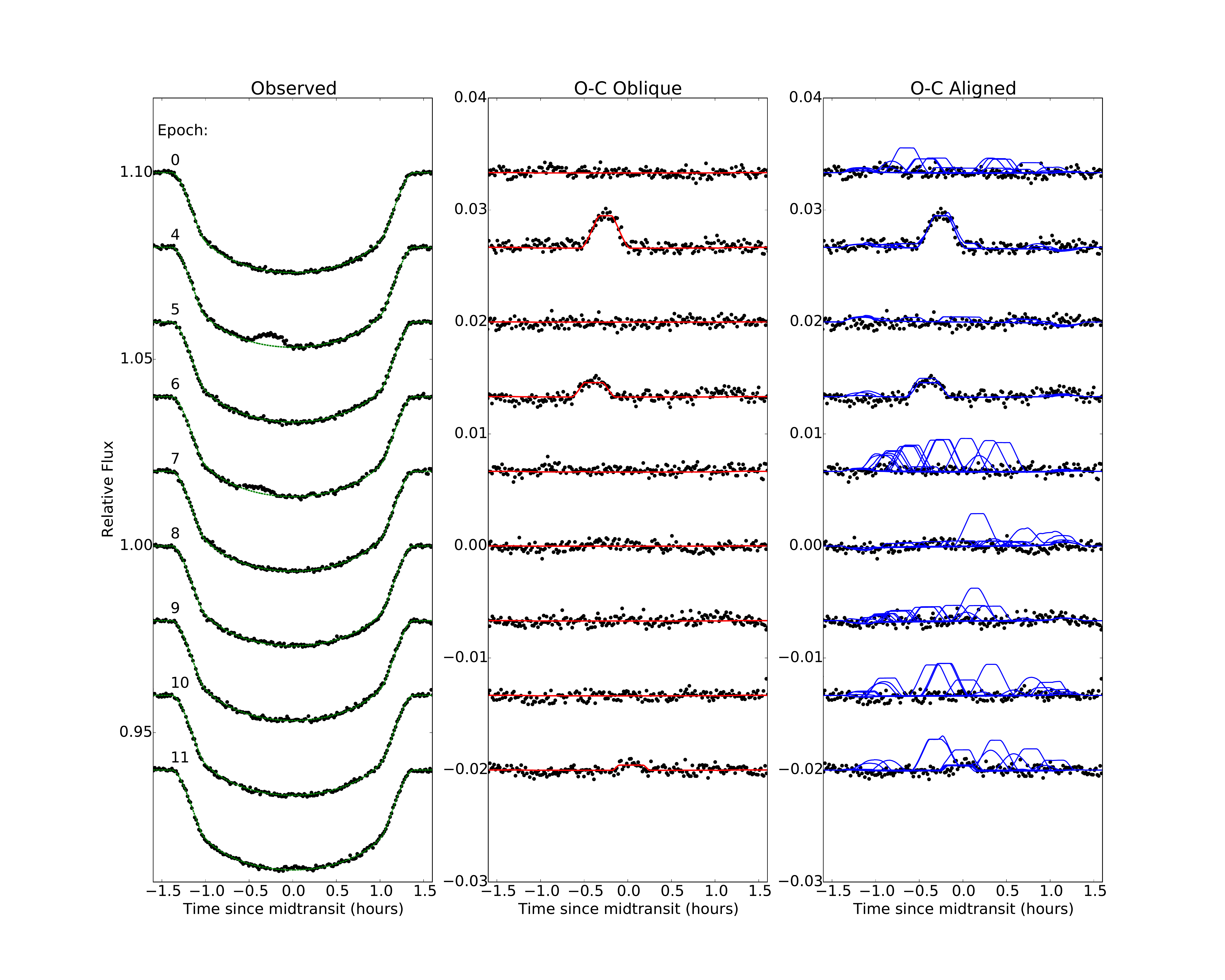}
\caption{{\bf Transits of WASP-107.}
    {\it Left.}---All the transits observed by {\it K2}. Epoch 1-3 are
  missing due to the loss of Module 4.  Vertical offsets were applied
  to separate the individual transits.  The dotted green line is the
  best-fitting model with no starspots.
  {\it Middle.}---Success of a high-obliquity model. Shown are
  residuals after subtracting the best-fitting no-spot model. The red
  curve is a representative model with obliquity $\Psi= 45^\circ$
  ($\lambda$ = 45$^{\circ}$, $i_{\star}$ = 90$^{\circ}$).
  {\it Right.}---Failure of low-obliquity models. The blue curves are
  20 representative models with $\Psi$<40$^{\circ}$. By construction,
  these models fit the observed the anomalies in Epoch 4, 6 and 11;
  but they also predict additional spot-crossing events that are not
  observed.}
\label{fig:combined}
\end{center}
\end{figure*}

\section{Stellar rotation period}
\label{rot}

Knowledge of the stellar rotation period is important for the analysis
of spot crossings.  The {\it K2} time series (Fig.~\ref{fig: lc})
shows variability at the 0.3\% level on a time scale of $\sim$10 days,
with what appear to be two cycles of a quasiperiodic function.  These
variations are characteristic of starspots being carried around by
stellar rotation. One estimate of the rotation period is the interval
between the two clearest minima, which we found to be $16.8 \pm
1.2$~days by fitting quadratic functions to the data surrounding each
minimum.  Another measure comes from the periodogram
\citep{Lomb1976,Scargle1982} (after masking out the transits), which
has a peak at $17.7^{+8.8}_{-2.8}$~days. Likewise the autocorrelation
method \citep{McQuillan2013} gives an estimate of
$17.0^{+2.1}_{-1.6}$~days. All these estimates agree with (and are
probably less accurate than) the period of $17.1 \pm 1.0$~days
reported by \citet{Anderson2017} based on two seasons of WASP
photometry. In what follows, we adopt $P_{\rm rot} = 17.1 \pm
1.0$~days.

The combination of $P_{\rm rot}$, $R_\star$, and $v\sin i_\star$ can be used
to calculate the stellar inclination angle $i_\star$:
\begin{equation}
  \sin i_\star = \frac{v\sin i_\star} {v} = \frac{v\sin i_\star}{2\pi R_\star/P_{\text{rot}}}.
\end{equation}
\citet{Anderson2017} reported $v\sin i_\star = 2.5 \pm
0.8$~km~s$^{-1}$ based on the observed broadening of the star's
spectral lines.  On the other hand, $v = 2\pi R_\star / P_{\rm rot} =
3.0 \pm 0.2$~km~s$^{-1}$ using the stellar radius in Tab. \ref{tab:para}. The comparison provides only a very loose
constraint on the orientation of the star: $\sin i_\star = 0.8 \pm
0.3$.

\section{Spot-Crossing Anomalies}
\label{sec: anomalies}

\begin{figure*}
\begin{center}
\includegraphics[width = 1.7\columnwidth]{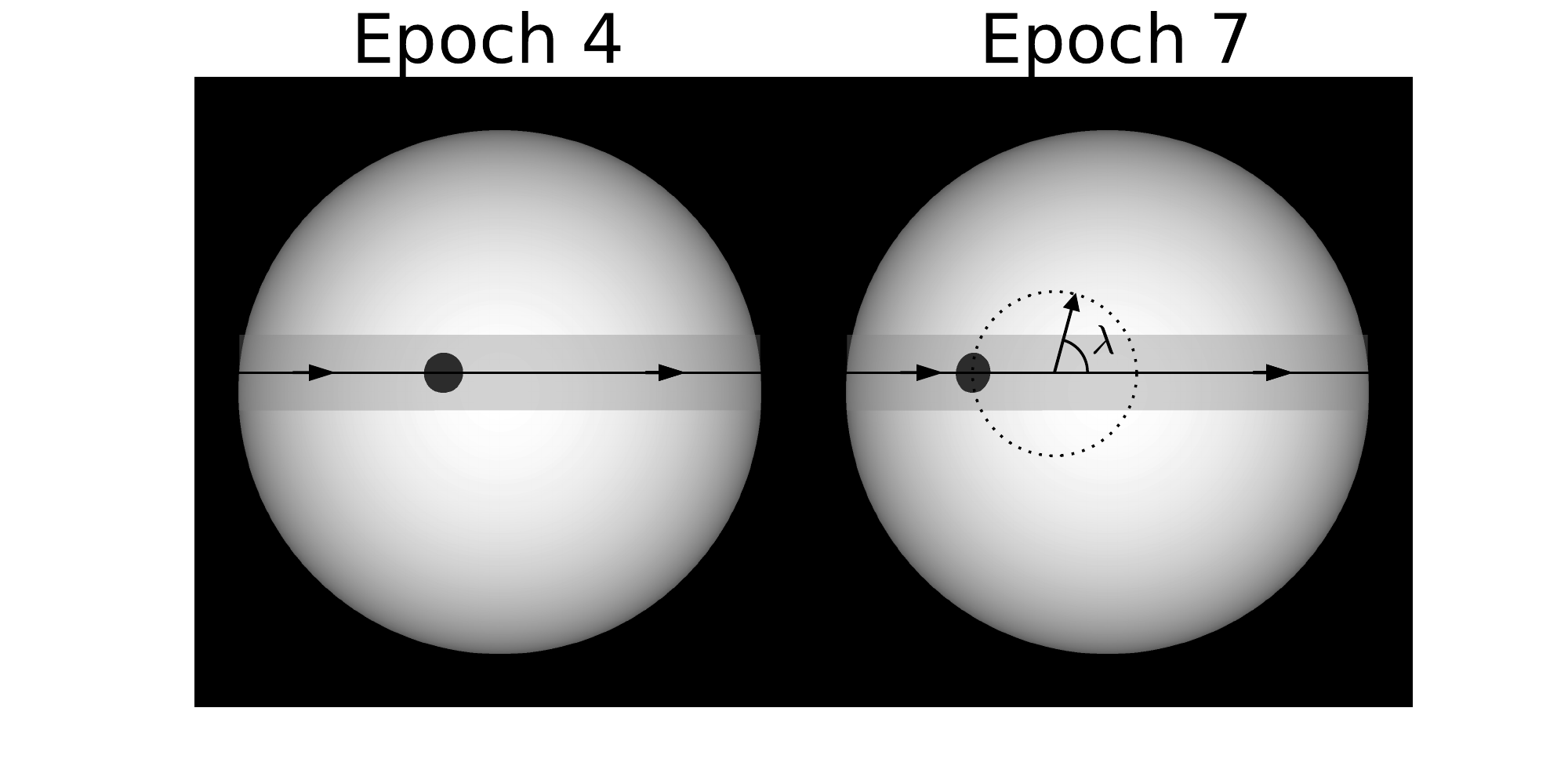}
\caption{{\bf Illustration of the geometric model.} {\it Left.}---The
  star during Epoch 4, when a spot-crossing anomaly was observed just
  before mid-transit. The dark circle is the starspot and the gray bar
  is the path of the transiting planet.  {\it Right.}---The predicted
  situation in Epoch 7 for zero obliquity and $P_{\rm
    rot}=18.1$~days. A spot-crossing anomaly should occur about
  one-quarter of the way through the transit, but is not observed. One
  way to avoid a spot crossing is to rotate the sky projection of the
  stellar rotation axis. By adjusting this angle $\lambda$ the
  predicted spot location can be changed to anywhere along the dotted
  line.}
\label{fig: skymodel}
\end{center}
\end{figure*}

At the time of an observed spot-crossing anomaly, the planet and the
spot overlap on the sky. Given the rotation period and the star's
orientation, we can calculate the spot's motion in the immediate
future and past. For obliquities near 0$^\circ$ or 180$^\circ$ spot
moves parallel to the transit chord, and is likely to be crossed
during the subsequent transits, or to have been crossed in the previous
transits. For higher obliquities this is unlikely because the spot's
path only intersects the transit chord at one location. The underlying
assumptions are that the spot lasts for at least one stellar
rotation, and that the spot is not much larger than the planet. These
assumptions seem justified here, given the apparent coherence of the
stellar variability over the $\approx$60 day observing interval, and
the small amplitude of the variations.

To determine the allowed range of obliquities for WASP-107 we used a
Monte Carlo procedure. In each of $10^4$ realizations, we drew values
of $P_{\rm rot}$, $\cos i_\star$, and the sky-projected obliquity
$\lambda$, from uniform distributions. We held the transit parameters
fixed in this part of the analysis, because they are so precisely
constrained.  Given the rotation period and stellar orientation, and
the observed times of the 3 anomalies, we calculated the trajectories
of each of the 3 spots over the entire time interval of the {\it K2}
observations. Then we computed a ``badness of fit'' statistic, by
totaling up the number of data points for which an intersection
between the planet and a spot was predicted but not observed.

Figure~\ref{fig: skymodel} illustrates the geometry. The left panel
shows the star at epoch 4 for the case $\lambda=0^\circ$ and
$i_\star=90^\circ$. The starspot's location along the transit chord is
fixed by the observed timing of the flux anomaly.  The right panel
shows the calculated location of the starspot at epoch 7.  According
to this model a flux anomaly should have been observed about
one-quarter of the way through the transit, but it was not.  The
dotted line shows all the possible locations for the starspot when
$\lambda$ is allowed to vary. For high enough $\lambda$, the spot can
avoid being crossed.  Similar plots can be made for the anomalies
observed in epochs 6 and 11.

The middle and right panels of Figure~\ref{fig:combined} 
illustrate the success of high-obliquity models and the failure of
low-obliquity models. The data points are the residuals after
subtracting the best-fitting transit model, allowing the spot-crossing
anomalies to be seen more clearly. In the middle panel, a good fit is
achieved by a model with true obliquity $\Psi = 45^{\circ}$. Of course, many other
high-obliquity models also fit the data. In the right panel we show 20
spot models drawn randomly from our Monte Carlo procedure, all having
$\Psi < 40^{\circ}$. None of them provide a satisfactory fit because
they predict anomalies when none are seen.

\begin{figure*}
\begin{center}

\includegraphics[width = 0.9\columnwidth]{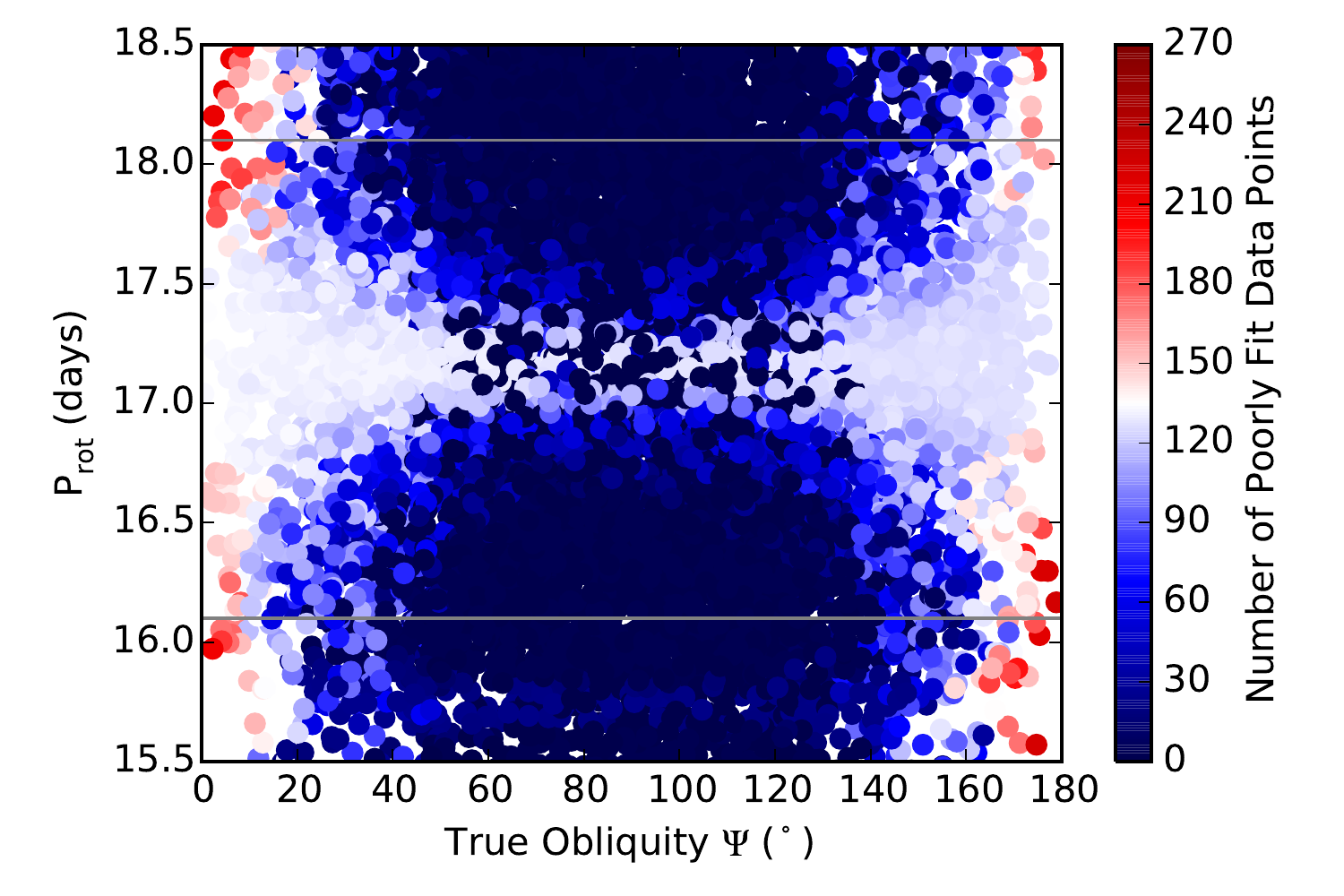}
\includegraphics[width = 0.9\columnwidth]{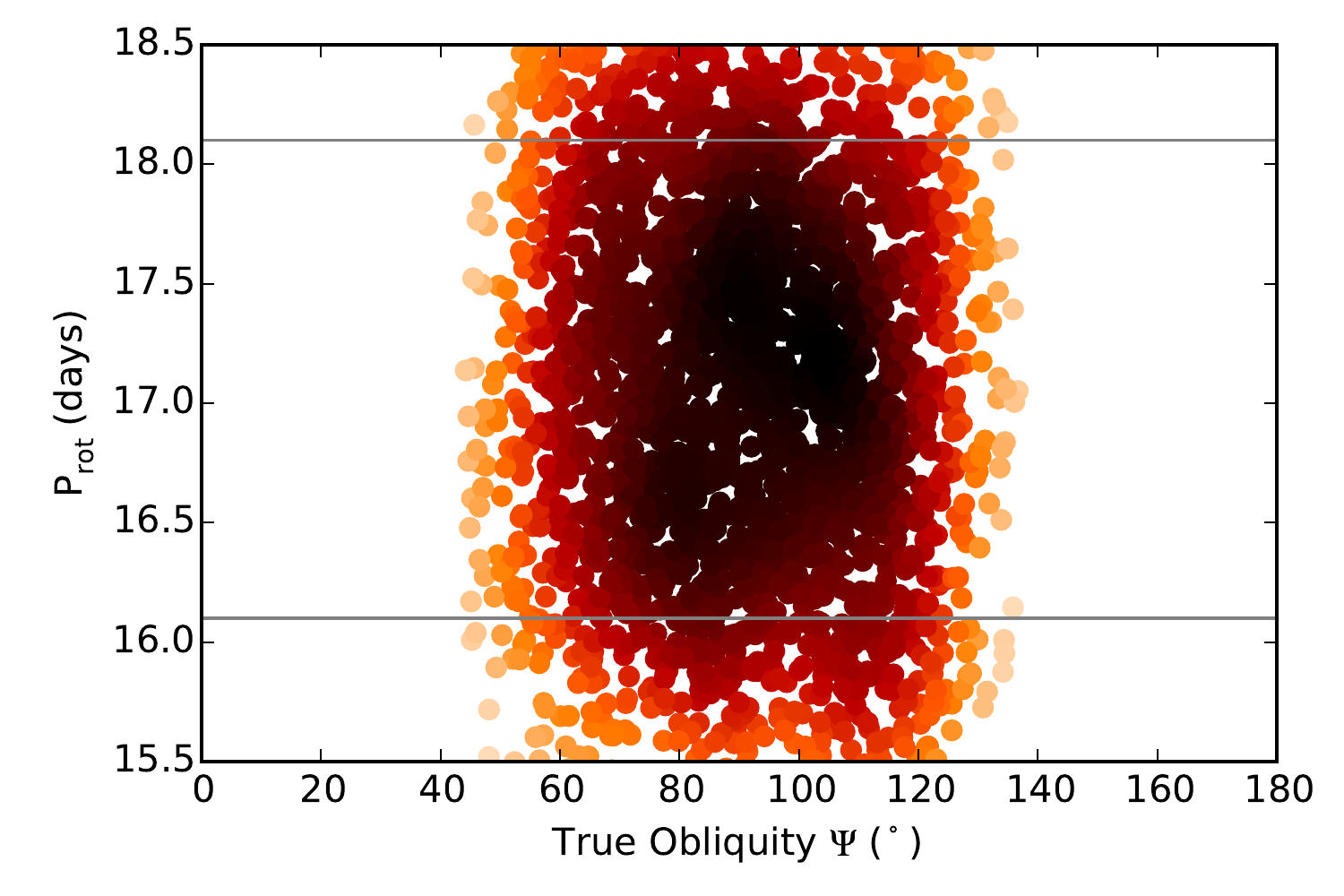}

\caption{{\bf Monte Carlo search for stellar orientations consistent
    with the data.} {\it Left.}---The quality of fit as a function of
  obliquity and rotation period. Horizontal gray lines bound the
  1$\sigma$ range in the measured rotation period. The color encodes
  the number of data points where anomalies are predicted but not
  observed. Dark blue indicates a good fit, and dark red a poor fit,
  with white in between.  {\it Right.}---Similar, but this time the
  color represents the density of good-fitting models in
  parameter space. In these models the true obliquity lies between
  about 40-140$^\circ$.}
\label{fig: geometry}
\end{center}
\end{figure*}

Figure~\ref{fig: geometry} summarizes the results. The left panel
shows the badness-of-fit, as a function of $P_{\rm rot}$ and the true
obliquity $\Psi$.  Two horizontal lines indicate the 1$\sigma$ range
of the measured rotation period.  All the best fits are obtained for
obliquities between 40-140$^\circ$.  Interestingly there is a narrow
range of rotation periods surrounding $17.16$ days for which the
obliquity must be restricted to a narrower range of $\Psi$ to
provide a good fit.  This is due to the nearly exact ratio of 3:1
between the stellar rotation period and the orbital period. When this
is the case, rotation brings the spot back to the transit chord just
in time to intersect the planet, regardless of the stellar
orientation. The right panel of Fig.~\ref{fig: geometry} shows the
density (in parameter space) of all the models that fit the data well,
and do not predict any unobserved anomalies. Here we see more clearly
that the successful models require the obliquity to be in the range
from about 40-140$^\circ$. Evidently the star is tilted at some large
angle, though we cannot specify the value of the angle with any
precision.

\section{Discussion}

Should there be any doubt about our spot analysis, the high obliquity
can be confirmed by detecting the Rossiter-McLaughlin (RM) effect
during transits. The expected signal amplitude is
\begin{equation}
\Delta V_{\rm{RM}} \approx 0.7 (R_p/R_\star)^2 v~\sin i_\star \approx 36~{\rm m~s}^{-1}.
\end{equation}
The situation brings to mind the case of HAT-P-11 \citep{Bakos2010},
in which the order of events was reversed.
\citep{Winn2010hat} found a high obliquity through RM
spectroscopy, but with a low signal-to-noise ratio; this finding was
subsequently confirmed by spot modeling
\citep{Sanchis-Ojeda2011Hat,Deming}. WASP-107 is about 2~mag fainter
than HAT-P-11, which will make the RM observation more challenging ---
but probably not impossible, given that the transit depth is also 5
times larger.

In fact these two systems form an interesting pair for
comparison. Both planets are ``super-Neptunes'' that orbit K dwarfs at
similar orbital distances. For WASP-107 and HAT-P-11, respectively,
the stellar masses are 0.69 and 0.80~$M_\odot$, the $a/R_\star$ values
are 18.1 and 15.1, and the planet masses are 1.5 and 2.2 times the
mass of Neptune.  And, both stars have high obliquities.  The one
prominent difference is that WASP-107b has a much larger diameter, and
is 9 times less dense.

\begin{figure*}
\begin{center}

\includegraphics[width = 2.\columnwidth]{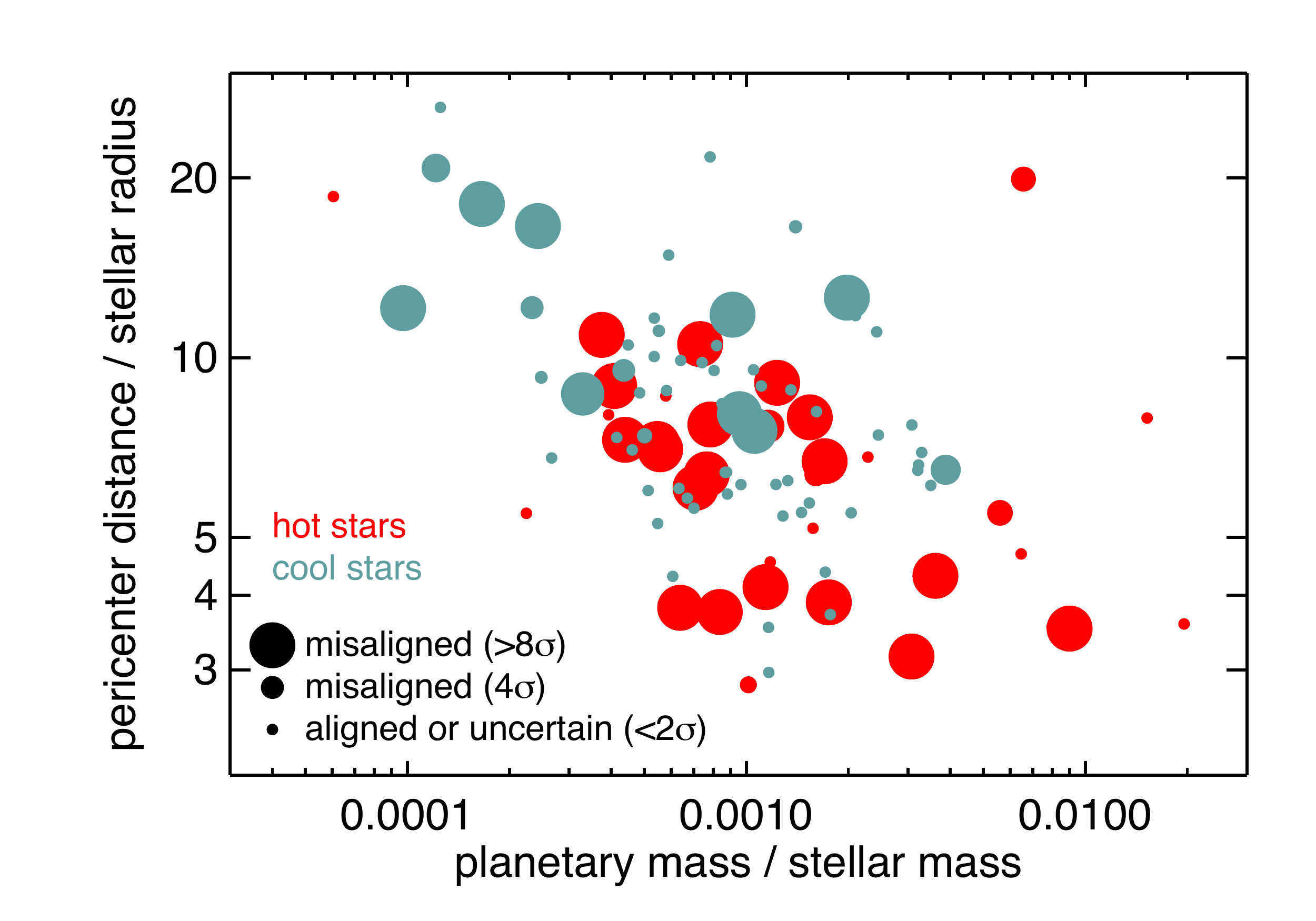}
\caption{{\bf Measurements of stellar obliquity} as a function of
  $M_p/M_\star$, the planet-to-star mass ratio, and $a_{\rm
    min}/R_\star$, the pericenter distance in units of the stellar
  radius.  Red points are for stars with $T_{\rm eff}>6200$~K
  (generally $>$1.2~$M_\odot$) and blue points are for cooler stars.
  The symbol size increases with $n_\sigma \equiv
  |\lambda|/\sigma_\lambda$, the confidence with which a nonzero
  obliquity can be excluded.  The smallest symbols have $n_\sigma<2$,
  the largest symbols have $n_\sigma>8$, and in between the symbol
  size is proportional to $n_\sigma$.  }
\label{fig:obliquity}
\end{center}
\end{figure*}

Both HAT-P-11 and WASP-107 are also exceptions to the general trend
that relatively low-mass stars ($M\lesssim1.2~M_\odot$ or $T_{\rm
  eff}\lesssim6200$~K) have low obliquities.  Fig.~\ref{fig:obliquity}
gathers together all the reliable obliquity measurements based on the
RM effect, spot modeling, or asteroseismology.\footnote{The data were
  collated with the help of
  \href{http://www.astro.keele.ac.uk/jkt/tepcat/tepcat.html}{\it
    TEPCat}, \href{http://www2.mps.mpg.de/homes/heller/}{Ren\'e
    Heller's webpage}, \href{http://exoplanet.eu/}{\tt exoplanet.eu}
  and \href{http://exoplanets.org}{\tt exoplanets.org}.  For those few
  systems with more than one planet, we used the properties of the
  most massive planet. We omitted 55~Cnc e, because although
  \citep{BourrierHebrard2014} reported a high obliquity, we are
  persuaded by the more precise data of \citep{LopezMorales+2014} that
  the obliquity is unknown.}  The size of each data point encodes our
confidence in each system's departure from perfect alignment; the
largest points are definitely misaligned. The data are shown as a
function of $M_p/M_\star$ and $a_{\rm min}/R_\star$, i.e., the
pericenter distance $a(1-e)$ in units of the stellar radius.  Red
points are for stars with $T_{\rm eff}>6200$~K, while blue points are
for cooler stars.

We call attention to two patterns: (1) for hot stars, the chance of
being significantly misaligned does not seem to depend on either of
these two parameters; (2) for cool stars, the misaligned stars are all
in the zone $a_{\rm min}/R_\star \gtrsim 8$.  Similar patterns were
noted earlier by \citep{Winn2010}, who speculated that the key
physical mechanism distinguishing these cases is tidal realignment. In
this picture, more rapid realignment is possible for cool stars,
because of more rapid dissipation associated with their thick
convective envelopes; and it is also more rapid for massive close-in
planets, because of the stronger tides they raise. Indeed
\citep{Albrecht2012} showed that for cool stars, the boundary between
aligned and misaligned stars could be expressed as a threshold value
of $(M_p/M_\star)^2(a/R_\star)^{-5}$, a parameter that should be
proportional to tidal dissipation rate in the theory of
\citet{Zahn1977}. Through modeling of the structure of each star,
\citep{ValsecchiRasio2014} even found evidence that the aligned stars
tend to have thicker convective zones.

However, tidal realignment theories suffer from a theoretical problem:
they need to avoid concomitant orbital decay
\citep{Winn2010,Dawson2014}. This problem might be superable
\citep{Lai2012,LiWinn2016}, but another problem emerged recently from
a study by \citep{Mazeh+2015}.  They found statistical evidence for
the hot/cool obliquity distinction even in cases where tidal
interactions should be utterly negligible, i.e., low-mass planets in
wide orbits. We also see from Fig.~\ref{fig:obliquity} that while high
$a_{\rm min}/R_\star$ is associated with misalignment, there is no
evidence for any separate dependence on $M_p/M_\star$, even though
this parameter should also be important in determining tidal
effects. For all these reasons, scenarios involving tidal realignment
are questionable.

Many scenarios have been presented to try and explain the high
obliquities of some planet-hosting stars. Any successful theory must
distinguish between hot and cool stars, and must also distinguish
between wide- and close-orbiting planets around cool stars. Mechanisms
that simply tilt the protoplanetary disk at an early stage, such as
the chaotic accretion of \citet{Bate+2010} or the stellar flybys of
\citet{Batygin2012}, do not have these properties. Neither do theories
involving planet scattering or the Kozai effect \citep[see,
  e.g.,][]{Fabrycky2007,Chatterjee2008}. These theories could be
combined with a tidal realignment mechanism to produce the hot/cool
distinction, but as we have already said, this solution is
problematic. \citet{Rogers+2012} proposed a mechanism to tilt hot
stars, but were silent about the cool stars with high obliquities.
\citet{MatsakosKonigl2015} and \citet{SpaldingBatygin2015} presented
two ways to avoid the theoretical problem with tidal realignment. The
former authors have the host star realign by ingesting a hot Jupiter
that is no longer observable, while the latter authors (following
\citet{Lai+2011}) invoke magnetic interactions with the inner edge of
the protoplanetary disk. However they do not say why these processes
should depend on the $a_{\rm min}/R_\star$ of the innermost surviving
giant planet. High priorities for future work are to verify that such
a dependence exists, by performing more measurements of systems with
large $a_{\rm min}/R_\star$, and then trying to explain why.

\bibliography{wasp107}

\begin{table}
\centering
\caption{System Parameters of WASP-107}
\begin{tabular}{lrr}
\hline
\hline
{\rm Parameter}  & {\rm }   & {\rm Ref.}  \\
\hline
    Stellar Parameters &$ 	    $&$   $\\
    $T_{\text{eff}} ~({\rm K})$ &$ 4430\pm 120     $& A \\
    $\log~g~(\text{dex})$ &$ 	4.5\pm 0.1  $& A \\
    $[\text{Fe/H}]~(\text{dex})$ &$ 	+0.02 \pm 0.10  $& A \\
    $v~\text{sin}~i_\star$ ~(km~s$^{-1})$ &$ 2.5 \pm 0.8   $& A \\
    $M_{\star} ~(M_{\odot})$ &$ 0.69 \pm 0.05     $& A \\
    $R_{\star} ~(R_{\odot})$ &$ 0.66 \pm 0.02     $& A \\
    $\text{Apparent $V$ mag}$ &$ 11.6     $& A \\
    $P_{\text{rot}} ~(\text{days})$ &$ 17.1 \pm 1.0     $& A \\
     $u_1$ &$ 0.6666 \pm 0.0062     $& B \\
     $u_2$ &$ 0.015 \pm 0.011     $& B \\
    \\
    Planetary Parameters &$ 	    $&$   $\\    

   $P~(\text{days})$ &$ 5.7214742 \pm 0.0000043    $& B \\
   $T_c~(\text{BJD})$ &$ 2457584.329897 \pm 0.000032    $& B \\
    $a/R_\star~$ &$   18.164 \pm 0.037 $& B \\
    $a~(\text{AU})$ &$   0.0558 \pm 0.0018  $& B \\
    $R_p/R_*~$ &$   0.14434 \pm 0.00018 $& B \\
    $R_p~(R_{\text{Jup}})$ &$    0.948 \pm 0.030 $& B \\
    $M_p~(M_{\text{Jup}})$ &$    0.12 \pm 0.01 $& B \\
    $i~(^{\circ})$ &$   89.8 \pm 0.2 $& B \\
    $b\equiv a\cos i/R_\star$ &$   0.07 \pm 0.07 $& B \\
    $e~({\rm assumed})$  &$  0    $& A \\

\hline
\end{tabular}
\tablecomments{A: \citet{Anderson2017}; B: this work.}
\label{tab:para}
\end{table}

\begin{table}
\centering
\caption{Spot-crossing anomalies observed in K2}
\begin{tabular}{lcccc}
\hline
\hline
Epoch  & $t_{\rm anom}$~(BJD~$-$~$2,454,833$) & Amplitude $A$ & Duration $\sigma_{\rm anom}$~(days) & Spot \# \\
\hline
$      4$ &$          2774.20512 \pm        0.00022    $& $              0.00298 \pm        0.00012   $& $              0.00493 \pm        0.00022    $ & 1 \\
$      6$ &$          2785.64242 \pm        0.00044    $& $              0.00156 \pm        0.00012   $& $              0.00493 \pm        0.00041    $ & 2 \\
$     11$ &$          2814.26773 \pm        0.00096    $& $              0.00064 \pm        0.00017   $& $              0.0031 \pm        0.0010    $ & 3 \\

\hline
\end{tabular}

\label{tab:anomalies}
\end{table}

\end{document}